\documentclass[fp,seceq]{jpsj3}
\usepackage{amsmath,amssymb,bm,graphicx}
\usepackage{xcolor}
\providecommand{\Tr}{\operatorname{Tr}}

\newcommand{\figpdf}[3][0.92\linewidth]{%
  \begin{figure}[tbp]
  \centering
    \IfFileExists{figs/#2.pdf}{%
      \makebox[\linewidth][c]{%
        \hspace*{-2mm}%
        \includegraphics[width=#1]{figs/#2.pdf}%
        \hspace*{2mm}%
      }%
    }{%
    \fbox{\begin{minipage}[c][34mm][c]{0.83\linewidth}
      \centering\bfseries Figure placeholder\\[3pt]
      \ttfamily\footnotesize\detokenize{figs/#2.pdf}\\[5pt]
      \normalfont\footnotesize Add the corresponding PDF in figs/.
    \end{minipage}}%
  }
  \caption{#3}\label{fig:#2}
  \end{figure}%
}

\title{Efficient Finite-Temperature Krylov Methods Based on Imaginary-Time Evolution}

\author{
Hiroto Yamamoto\thanks{E-mail: yamamoto.hiroto.r8@dc.tohoku.ac.jp}
and
Katsuhiro Morita\thanks{E-mail: katsuhiro.morita.c7@mso.tohoku.ac.jp}
}

\inst{
Department of Physics, Graduate School of Science,
Tohoku University, Sendai 980-8578, Japan
}

\abst{
We develop two finite-temperature Krylov methods
that construct Krylov subspaces using imaginary-time
evolution rather than direct applications of the Hamiltonian.
The finite-temperature imaginary-time Lanczos method (FTiTL)
applies the Lanczos method to the imaginary-time evolution operator,
whereas the finite-temperature imaginary-time Krylov method (FTiTK)
constructs a generalized eigenvalue problem
using a nonorthogonal Krylov basis.
We evaluate the accuracy and computational efficiency
of these methods for the square-lattice transverse-field
Ising model, using exact diagonalization for $N=16$
and quantum Monte Carlo results for $N=32$ as references.
With appropriately chosen imaginary-time steps,
both methods achieve higher accuracy than the conventional
finite-temperature Lanczos method (FTLM)
with fewer local operator applications
in the ferromagnetic regime and near the quantum critical point.
For $N=32$, FTiTL requires approximately half as many
local operator applications as FTLM to achieve
a root-mean-square error of $10^{-2}$ or less
in the specific heat $C/N$.
Furthermore, FTiTK can be implemented using only one state vector
of full Hilbert-space dimension, halving the memory required
for storing state vectors relative to FTLM.
}

\begin{document}
\maketitle

\section{Introduction}\label{sec:intro}

In finite-temperature calculations of quantum many-body systems,
the exponential growth of the Hilbert space limits the applicability
of exact diagonalization to small systems.
Although quantum Monte Carlo (QMC) methods can be applied to larger systems,
calculations become difficult when the sign problem arises,
as in frustrated systems.\cite{SandvikKurkijarvi1991,Evertz2003,TodoKato2001,HeneliusSandvik2000,Troyer2005}
The finite-temperature Lanczos method (FTLM)
combines stochastic trace estimation with the Lanczos method
to approximate thermodynamic quantities.\cite{Jaklic1994,PrelovsekBonca2013,LiLin2022}
FTLM is free from the sign problem
and has become a useful approach for frustrated quantum spin systems,
including kagome-lattice antiferromagnets.\cite{Schnack2018,Morita2022}
However, FTLM is subject to statistical errors
and errors due to the finite dimension of the Krylov subspace,
and improved variants have been investigated to enhance its accuracy.
\cite{Schnack2020,Aichhorn2003,MoritaTohyama2020}
Krylov methods based on imaginary-time evolution
have also been used to calculate ground and excited states.
\cite{Motta2020,YeterAydeniz2020}
Finite-temperature methods for quantum computers
that combine Krylov subspaces generated by real-time evolution
with stochastic trace estimation have also been proposed.
\cite{Gentinetta2026,YamamotoMorita2026}
In addition, finite-temperature methods based on quantum typicality,
using thermal pure quantum (TPQ) states,
have been proposed.\cite{Sugiura2012,Sugiura2013}

In this study, we investigate improvements in computational efficiency
over conventional FTLM by using Krylov subspaces generated
by the imaginary-time evolution operator
$B=\mathrm{e}^{-\eta H}$ ($\eta>0$)
for finite-temperature stochastic trace estimation
on classical computers.
Specifically, we consider the finite-temperature imaginary-time
Lanczos method (FTiTL), which applies the Lanczos method to $B$,
and the finite-temperature imaginary-time Krylov method (FTiTK),
which constructs a generalized eigenvalue problem
from a nonorthogonal Krylov basis.
Although both methods use the same Krylov subspace,
they treat it differently.
In particular, FTiTK can be implemented using only one state vector
of full Hilbert-space dimension,
reducing the memory required for storing state vectors
to half that of FTLM.
We apply these methods to the square-lattice transverse-field Ising model.
For the 16-site system, we compare their accuracy
with that of conventional FTLM and the canonical TPQ (c-TPQ) method,
using exact diagonalization as a reference.
For the 32-site system, we use QMC results as a reference
to compare the accuracy and number of local operator applications
with those of FTLM.
We also evaluate errors associated with the Trotter approximation
and differences in memory requirements.

In Sec.~\ref{sec:methods}, we formulate the three Krylov methods.
In Sec.~\ref{sec:setup}, we introduce the model
and define the computational cost.
Numerical results are presented in Sec.~\ref{sec:results},
and the characteristics and applicability of the methods
are discussed in Sec.~\ref{sec:discussion}.

\section{Methods}
\label{sec:methods}
Whereas FTLM constructs Krylov subspaces from the Hamiltonian $H$,
FTiTL and FTiTK construct Krylov subspaces using
the imaginary-time evolution operator $B=\mathrm{e}^{-\eta H}$.
Because imaginary-time evolution exponentially suppresses
high-energy components, it is expected to improve
the convergence of Ritz values associated with low-energy states,
which make important contributions to finite-temperature
thermodynamic quantities.

\subsection{Conventional finite-temperature Lanczos method}
\label{sec:ftlm}

We set the Boltzmann constant to $k_{\mathrm B}=1$
and define the inverse temperature as $\beta=1/T$.
The partition function $Z$, energy $E$,
specific heat $C$, and entropy $S$
for the Hamiltonian $H$ are given by
\begin{align}
 Z(\beta)&=\Tr\mathrm{e}^{-\beta H},
 \label{eq:Z}\\
 E(\beta)&=\langle H\rangle_\beta,
 \label{eq:energy}\\
 C(\beta)&=\beta^2
 \left(\langle H^2\rangle_\beta-\langle H\rangle_\beta^2\right),
 \label{eq:heat}\\
 S(\beta)&=\ln Z(\beta)+\beta E(\beta)
 \label{eq:entropy}
\end{align}
where $\langle A\rangle_\beta
=\Tr(A\mathrm{e}^{-\beta H})/Z(\beta)$ denotes
the thermal expectation value of an operator $A$.

In FTLM, the trace of an operator $A$ is approximated
using $R$ normalized random states $|r\rangle$
distributed isotropically in the Hilbert space:
\begin{equation}
 \Tr A\simeq
 \frac{\dim\mathcal{H}}{R}
 \sum_{r=1}^{R}\langle r|A|r\rangle
 \label{eq:trace}
\end{equation}
Quantum typicality allows thermal averages
to be approximated accurately with only a small number
of random states at high temperatures.
\cite{Hams2000,Reimann2007,Sugiura2012,Sugiura2013,Steinigeweg2014}
Up to an observable-dependent prefactor,
the relative statistical error in the thermal average
is expected to scale as follows~\cite{Schnack2020}:
\begin{equation}
\frac{\delta\langle A\rangle_\beta}
{|\langle A\rangle_\beta|}
=O\!\left(
\frac{1}{\sqrt{RZ_{\mathrm{eff}}(\beta)}}
\right).
\label{eq:typicality_error}
\end{equation}
Here, $\delta\langle A\rangle_\beta$ denotes
the standard deviation of the estimate,
$Z_{\mathrm{eff}}(\beta)
=\Tr\mathrm{e}^{-\beta(H-E_0)}$,
and $E_0$ is the ground-state energy.
The quantity $Z_{\mathrm{eff}}(\beta)$ provides
an estimate of the effective number of thermally contributing states.

For each random state $|r\rangle$, FTLM further constructs
the Krylov subspace
\begin{equation}
 \mathcal{K}_D(H,|r\rangle)
 =\operatorname{span}
 \{|r\rangle,H|r\rangle,\ldots,H^{D-1}|r\rangle\}
 \label{eq:K_H}
\end{equation}
Using the orthonormal basis
$\{|q_{n,\mathrm{FTLM}}^{(r)}\rangle\}_{n=0}^{D-1}$
generated by the Lanczos method,\cite{Lanczos1950}
projecting $H$ onto this subspace yields
a $D\times D$ tridiagonal matrix
$T_{\mathrm{FTLM}}^{(r)}$.
Solving the eigenvalue problem
\begin{equation}
 T_{\mathrm{FTLM}}^{(r)}\bm v_j^{(r)}
 =
 \varepsilon_{j,\mathrm{FTLM}}^{(r)}\bm v_j^{(r)}
 \label{eq:ftlm_eigenproblem}
\end{equation}
gives the Ritz energies
$\varepsilon_{j,\mathrm{FTLM}}^{(r)}$.
The corresponding Ritz states are expressed as
$|\psi_{j,\mathrm{FTLM}}^{(r)}\rangle
=\sum_{n=0}^{D-1}(v_j^{(r)})_n
|q_{n,\mathrm{FTLM}}^{(r)}\rangle$,
where $(v_j^{(r)})_n$ denotes the $n$th component
of the normalized eigenvector $\bm v_j^{(r)}$.

The weights associated with these Ritz states are defined as
$w_{j,\mathrm{FTLM}}^{(r)}
=|\langle\psi_{j,\mathrm{FTLM}}^{(r)}|r\rangle|^2
=|(v_j^{(r)})_0|^2$.
The contribution of each random state
to the partition function is then approximated by
\begin{equation}
 \langle r|\mathrm{e}^{-\beta H}|r\rangle
 \simeq
 \sum_{j=0}^{D-1}
 w_{j,\mathrm{FTLM}}^{(r)}
 \mathrm{e}^{-\beta\varepsilon_{j,\mathrm{FTLM}}^{(r)}}
 \label{eq:ftlmZ}
\end{equation}
Thus, from Eqs.~(\ref{eq:trace}) and (\ref{eq:ftlmZ}),
the FTLM approximation to the partition function is
\begin{equation}
 Z_{\mathrm{FTLM}}(\beta)
 =\frac{\dim\mathcal{H}}{R}
 \sum_{r=1}^{R}\sum_{j=0}^{D-1}
 w_{j,\mathrm{FTLM}}^{(r)}
 \mathrm{e}^{-\beta\varepsilon_{j,\mathrm{FTLM}}^{(r)}}
 \label{eq:ftlm_final_partition}
\end{equation}
Similarly, $\langle H\rangle_\beta$ and
$\langle H^2\rangle_\beta$ can be evaluated
by multiplying each term by the corresponding
Ritz energy or its square.
When constructing the tridiagonal matrix
$T_{\mathrm{FTLM}}^{(r)}$,
the storage of the preceding Lanczos vector
can be reused while sequentially accumulating
the contributions from the local terms of the Hamiltonian.
Thus, the Lanczos iterations in FTLM can be performed
using two state vectors.

In the following, we distinguish the three methods by
$X\in\{\mathrm{FTLM},\mathrm{FTiTL},\mathrm{FTiTK}\}$
and denote their Ritz states, energies, and weights by
$|\psi_{j,X}^{(r)}\rangle$,
$\varepsilon_{j,X}^{(r)}$,
and $w_{j,X}^{(r)}$, respectively.
The partition function for each method
can then be written in the common form
\begin{equation}
 Z_X(\beta)=\frac{\dim\mathcal{H}}{R}
 \sum_{r=1}^{R}\sum_{j=0}^{D-1}
 w_{j,X}^{(r)}
 \mathrm{e}^{-\beta\varepsilon_{j,X}^{(r)}}
 \label{eq:Z_common}
\end{equation}

\subsection{Finite-temperature imaginary-time Lanczos method}
\label{sec:exp}

Unlike FTLM, which uses $H$ to generate Krylov subspaces,
the two imaginary-time methods use the operator
\begin{equation}
 B=\mathrm{e}^{-\eta H},\qquad
 \eta=\frac{\rho}{W}>0
 \label{eq:B}
\end{equation}
Here, $W$ is the spectral width of $H$,
and $\rho>0$ is a dimensionless parameter
that controls the imaginary-time step.
Let $|E_i\rangle$ denote an eigenstate of $H$
with the corresponding eigenenergy $E_i$.
Then,
\[
H|E_i\rangle=E_i|E_i\rangle,\qquad
B|E_i\rangle=\mathrm{e}^{-\eta E_i}|E_i\rangle
\]
Thus, $H$ and $B$ share the same eigenstates,
and the eigenvalues of $B$ are given by $\mathrm{e}^{-\eta E_i}$.

In FTiTL, the Lanczos method is applied
to the Krylov subspace
\begin{equation}
 \mathcal{K}_D(B,|r\rangle)
 =\operatorname{span}
 \{|r\rangle,B|r\rangle,\ldots,B^{D-1}|r\rangle\}
 \label{eq:K_B}
\end{equation}
Projecting $B$ onto this subspace
using the orthonormal basis
$\{|q_{n,\mathrm{FTiTL}}^{(r)}\rangle\}_{n=0}^{D-1}$
yields a $D\times D$ tridiagonal matrix
$T_{\mathrm{FTiTL}}^{(r)}$.
The corresponding eigenvalue problem is
\begin{equation}
 T_{\mathrm{FTiTL}}^{(r)}\bm v_j^{(r)}
 =\lambda_j^{(r)}\bm v_j^{(r)}
 \label{eq:exp_eigenproblem}
\end{equation}
where $\lambda_j^{(r)}$ denotes a Ritz value of $B$.
The corresponding Ritz state is given by
$|\psi_{j,\mathrm{FTiTL}}^{(r)}\rangle
=\sum_{n=0}^{D-1}(v_j^{(r)})_n
|q_{n,\mathrm{FTiTL}}^{(r)}\rangle$.
The energies and weights are then obtained
from the Ritz values and states as
\begin{align}
 \varepsilon_{j,\mathrm{FTiTL}}^{(r)}
 &=-\frac{1}{\eta}\ln\lambda_j^{(r)},
 \label{eq:inverse}\\
 w_{j,\mathrm{FTiTL}}^{(r)}
 &=|\langle\psi_{j,\mathrm{FTiTL}}^{(r)}|r\rangle|^2
 =|(v_j^{(r)})_0|^2
 \label{eq:exp_weight}
\end{align}
In FTiTL, the partition function is evaluated
by substituting $\varepsilon_{j,\mathrm{FTiTL}}^{(r)}$
and $w_{j,\mathrm{FTiTL}}^{(r)}$
into Eq.~(\ref{eq:Z_common}).
As in conventional FTLM, errors arise due to
the finite dimension of the Krylov subspace.
In practical calculations, additional errors
arise from the Trotter approximation.

When constructing the tridiagonal matrix
$T_{\mathrm{FTiTL}}^{(r)}$,
FTiTL requires the application of
the Trotter-approximated imaginary-time evolution operator.
This operation requires a workspace vector
in addition to the two state vectors
used in the Lanczos iterations.
Therefore, our implementation of FTiTL
stores a total of three state vectors.

\subsection{Finite-temperature imaginary-time Krylov method}
\label{sec:iftqk}

FTiTK uses the same Krylov subspace
as FTiTL in Eq.~(\ref{eq:K_B})
but retains a nonorthogonal basis.
Unlike the Lanczos method,
FTiTK does not orthonormalize the basis
but instead constructs a generalized eigenvalue problem
from the overlaps between basis vectors.

For each random state $|r\rangle$, we define
\begin{equation}
 \mu_n^{(r)}=\langle r|B^n|r\rangle,\qquad
 |\phi_n^{(r)}\rangle=B^n|r\rangle
 \label{eq:moments}
\end{equation}
Here, $\mu_n^{(r)}$ denotes the overlap
with the initial state, referred to as a moment.
Projecting $B$ onto the nonorthogonal basis
$\{|\phi_n^{(r)}\rangle\}_{n=0}^{D-1}$
leads to the generalized eigenvalue problem
\begin{equation}
 \tilde{B}^{(r)}\bm v_j^{(r)}
 =\lambda_j^{(r)}\tilde{S}^{(r)}\bm v_j^{(r)}
\label{eq:gevp}
\end{equation}
where $\tilde{B}^{(r)}$ is the projected matrix,
$\tilde{S}^{(r)}$ is the overlap matrix,
and $\lambda_j^{(r)}$ denotes a Ritz value of $B$.
Because $B$ is Hermitian,
the matrix elements can be expressed as
\begin{equation}
 \begin{aligned}
 \tilde{S}_{ij}^{(r)}
 &=\langle\phi_i^{(r)}|\phi_j^{(r)}\rangle
 =\mu_{i+j}^{(r)},\\
 \tilde{B}_{ij}^{(r)}
 &=\langle\phi_i^{(r)}|B|\phi_j^{(r)}\rangle
 =\mu_{i+j+1}^{(r)}
 \end{aligned}
 \label{eq:hankel}
\end{equation}
Thus, both matrices can be constructed
from the $2D$ moments
$\mu_0^{(r)},\ldots,\mu_{2D-1}^{(r)}$.

We normalize the generalized eigenvectors
$\bm v_j^{(r)}$ such that
$(\bm v_j^{(r)})^\dagger
\tilde{S}^{(r)}\bm v_k^{(r)}=\delta_{jk}$.
The corresponding Ritz states are then given by
$|\psi_{j,\mathrm{FTiTK}}^{(r)}\rangle
=\sum_{n=0}^{D-1}(v_j^{(r)})_n|\phi_n^{(r)}\rangle$.
The energies and weights are then obtained as
\begin{align}
 \varepsilon_{j,\mathrm{FTiTK}}^{(r)}
 &=-\frac{1}{\eta}\ln\lambda_j^{(r)},
 \label{eq:ftqk_energy}\\
 w_{j,\mathrm{FTiTK}}^{(r)}
 &=|\langle\psi_{j,\mathrm{FTiTK}}^{(r)}|r\rangle|^2
 =\left|
 \sum_{n=0}^{D-1}(v_j^{(r)})_n\mu_n^{(r)}
 \right|^2
 \label{eq:weight}
\end{align}
Substituting these quantities into Eq.~(\ref{eq:Z_common})
gives the FTiTK approximation to the partition function.

The overlap matrix $\tilde{S}^{(r)}$ may become
ill-conditioned due to linear dependence
among the nonorthogonal basis vectors
or numerical errors.
To address this issue, we regularize
the generalized eigenvalue problem by discarding
directions corresponding to eigenvalues
$s_i<10^{-13}$ of $\tilde{S}^{(r)}$
and solving the problem in the remaining subspace.
\cite{Epperly2022,Klymko2022,YamamotoMorita2026}
Although FTiTL and FTiTK yield identical Ritz information
in exact arithmetic,
the accuracy of FTiTK may be degraded
by numerical errors associated with the nonorthogonal basis
and by regularization.

We next describe how to compute
$\mu_0^{(r)},\ldots,\mu_{2D-1}^{(r)}$
sequentially using a single state vector.
For the actual imaginary-time evolution,
we use the symmetric Trotter approximation $B_{\rm T}$,
which is factorized as
\begin{equation}
 B_{\rm T}=G^\dagger G
 \label{eq:Gfact}
\end{equation}
The explicit construction of $G$
is described in Sec.~\ref{sec:cluster}.
In the following, we replace $B$ with $B_{\rm T}$
and use the same notation $\mu_n^{(r)}$
for the corresponding moments.

Starting from $|a_0^{(r)}\rangle=|r\rangle$,
we update the state for $k\ge0$ according to
\begin{equation}
 \begin{aligned}
 |a_{2k+1}^{(r)}\rangle
 &=G|a_{2k}^{(r)}\rangle,\\
 |a_{2k+2}^{(r)}\rangle
 &=G^\dagger|a_{2k+1}^{(r)}\rangle
 \end{aligned}
 \label{eq:alternating}
\end{equation}
The moments then satisfy
\begin{equation}
 \mu_{2k}^{(r)}
 =\|a_{2k}^{(r)}\|^2,\qquad
 \mu_{2k+1}^{(r)}
 =\|a_{2k+1}^{(r)}\|^2
 \label{eq:norm_mom}
\end{equation}
Thus, the required moments can be obtained
simply by recording the squared norm after each update.
Computing $\mu_0^{(r)},\ldots,\mu_{2D-1}^{(r)}$
requires $D$ applications of $G$
and $D-1$ applications of $G^\dagger$.
Because this procedure does not require
the initial state to be stored separately,
the moments can be generated using a single state vector
by applying $G$ and $G^\dagger$ in-place.

In summary, our implementation requires
two full Hilbert-space state vectors for FTLM
and three for FTiTL,
whereas FTiTK requires only one.
Consequently, compared with FTLM,
FTiTL requires 1.5 times as much memory
for storing state vectors,
while FTiTK reduces this memory requirement by half.

\section{Model}\label{sec:setup}
\subsection{Hamiltonian and Trotter decomposition}
\label{sec:cluster}

We consider the two-dimensional square-lattice transverse-field
Ising model with $N$ sites.
The Hamiltonian is expressed in terms of
the Pauli matrices $\sigma_i^x,\sigma_i^z$ as
\begin{equation}
 H=-J\sum_{\langle i,j\rangle}\sigma_i^z\sigma_j^z
   -h\sum_{i=1}^{N}\sigma_i^x
 \label{eq:tfim}
\end{equation}
where $\langle i,j\rangle$ denotes a pair of nearest-neighbor sites,
$J>0$ is the strength of the ferromagnetic Ising interaction,
and $h$ is the transverse-field strength.
We set $J=1$ throughout this study.

To apply the Trotter decomposition to the imaginary-time
evolution operator, we divide the Hamiltonian
into the interaction and transverse-field terms:
\begin{equation}
 H_{ZZ}=-J\sum_{\langle i,j\rangle}\sigma_i^z\sigma_j^z,
 \qquad H_X=-h\sum_i\sigma_i^x
 \label{eq:splitH}
\end{equation}
Because these terms generally do not commute,
we approximate the imaginary-time evolution operator $B$
using the symmetric second-order Suzuki--Trotter decomposition
\cite{Trotter1959,Suzuki1976}
\begin{equation}
 B_{\rm T}=
 \left[
 \mathrm{e}^{-\delta H_{ZZ}/2}
 \mathrm{e}^{-\delta H_X}
 \mathrm{e}^{-\delta H_{ZZ}/2}
 \right]^M,
 \qquad \delta=\frac{\eta}{M}
 \label{eq:trotter}
\end{equation}
where $M$ is the number of Trotter steps.
This decomposition preserves the Hermiticity of $B_{\rm T}$.
In particular, for $M=1$, the operator $G$
in Eq.~(\ref{eq:Gfact}) can be defined as
\begin{equation}
 G=\mathrm{e}^{-\eta H_X/2}\mathrm{e}^{-\eta H_{ZZ}/2}
 \label{eq:G}
\end{equation}

For the numerical calculations, we use 16- and 32-site
square-lattice clusters with periodic boundary conditions (PBC).
For $N=16$, we employ a $4\times4$ square lattice
with periodicity vectors
$\bm T_1=(4,0)$ and $\bm T_2=(0,4)$.
For $N=32$, we use a cluster defined by
the periodicity vectors
$\bm T_1=(4,4)$ and $\bm T_2=(-4,4)$.
Both clusters have 90-degree rotational symmetry.
We consider two transverse-field strengths:
$h=1$ in the ferromagnetic regime
and the quantum critical field
$h/J\simeq3.04438$\cite{Blote2002}
in the thermodynamic limit.

\subsection{Local operator-action counts}
\label{sec:cost}

In the Krylov methods considered here,
repeated evaluations of $H|\psi\rangle$ and $B_{\rm T}|\psi\rangle$
constitute the dominant computational cost.
To compare the computational costs of the three methods
while minimizing dependence on code implementation
and computing environment,
we use the total number of local operator applications
$L_X$ required to perform these operations as a cost measure.
We count each application of
$\sigma_i^z\sigma_j^z$, $e^{a\sigma_i^z\sigma_j^z}$,
$\sigma_i^x$, and $e^{b\sigma_i^x}$
as one local operator application.

On the square lattice, there are $2N$ nearest-neighbor bonds
and $N$ transverse-field terms.
Thus, a single application of $H$ requires
$3N$ local operator applications.
The total number of local operator applications for FTLM is therefore
\begin{equation}
 L_{\mathrm{FTLM}}=3NRD_{\mathrm{FTLM}}
 \label{eq:WH}
\end{equation}
where $D_{\mathrm{FTLM}}$ is the Krylov dimension.
When constructing the projected matrix
of dimension $D_{\mathrm{FTLM}}$,
an application of the Hamiltonian is also required
to evaluate the final diagonal element.
We therefore count $D_{\mathrm{FTLM}}$
Hamiltonian applications.

In FTiTL, combining adjacent $ZZ$ half-steps
in the symmetric Trotter decomposition of Eq.~(\ref{eq:trotter})
reduces a single application of $B_{\rm T}$
to $M+1$ $ZZ$ layers and $M$ $X$ layers.
Each layer requires $2N$ and $N$
local operator applications, respectively.
Thus,
\begin{equation}
 L_{\mathrm{FTiTL}}
 =NRD_{\mathrm{FTiTL}}(3M+2)
 \label{eq:WB}
\end{equation}
where $D_{\mathrm{FTiTL}}$ is the Krylov dimension.

In FTiTK, constructing the generalized eigenvalue problem
of dimension $D_{\mathrm{FTiTK}}$
requires the moments
$\mu_0,\ldots,\mu_{2D_{\mathrm{FTiTK}}-1}$.
According to Eq.~(\ref{eq:alternating}),
this requires $D_{\mathrm{FTiTK}}$ applications of $G$
and $D_{\mathrm{FTiTK}}-1$ applications of $G^\dagger$.
Splitting the symmetric Trotter product at its center,
each application of $G$ or $G^\dagger$ requires
$\lfloor M/2\rfloor+1$ $ZZ$ layers
and $\lceil M/2\rceil$ $X$ layers.
Consequently,
\begin{equation}
 L_{\mathrm{FTiTK}}
 =NR(2D_{\mathrm{FTiTK}}-1)
 \left(M+\left\lfloor\frac{M}{2}\right\rfloor+2\right)
 \label{eq:WQ}
\end{equation}
In particular, for $M=1$,
which is the main computational setting in this study,
the local operator-action counts for the three methods become
\begin{equation}
 \begin{aligned}
 \frac{L_{\mathrm{FTLM}}}{NR}
 &=3D_{\mathrm{FTLM}},\\
 \frac{L_{\mathrm{FTiTL}}}{NR}
 &=5D_{\mathrm{FTiTL}},\\
 \frac{L_{\mathrm{FTiTK}}}{NR}
 &=3(2D_{\mathrm{FTiTK}}-1)
 \end{aligned}
\end{equation}
At the same Krylov dimension,
the imaginary-time methods require more local operator applications
than FTLM.
However, if the desired accuracy can be achieved
with a smaller Krylov dimension,
the total number of local operator applications can be reduced.

We note that this cost measure does not account
for differences in the computational costs
of diagonal and off-diagonal terms
or the memory traffic associated with each operation,
as these depend on the code implementation
and computing environment.
Therefore, $L_X$ does not directly represent
the actual computation time.

\section{Numerical Results}\label{sec:results}

In this section, we examine the numerical accuracy of FTLM, FTiTL, and FTiTK.
First, we compare the Krylov methods and the c-TPQ method
with exact diagonalization (ED) for the $N=16$ system
and examine their convergence.
Next, for the $N=32$ system, we evaluate
the temperature dependence of the specific heat and entropy
using quantum Monte Carlo (QMC) results as reference values.
For the QMC calculations, we employ the loop algorithm
implemented in ALPS, based on the stochastic series expansion (SSE).
\cite{Alet2026,ALPS2011,EvertzLana1993,BeardWiese1996,Evertz2003,TodoKato2001}
The SSE method does not require the Trotter approximation
and is therefore free from systematic errors
associated with imaginary-time discretization.
The entropy is calculated by thermodynamic integration
using the energy obtained from QMC.
The statistical errors in the QMC specific heat $C/N$
used as reference values are approximately
$3\times10^{-4}$ for $0.1\leq T/J\leq 0.3$
and below $10^{-4}$ at the other temperature points.

The initial random states for each Krylov method are generated as
\begin{equation}
 |r\rangle
 =\frac{1}{\sqrt{\mathcal N_r}}
 \sum_{n=0}^{\dim\mathcal{H}-1}
 g_n^{(r)}|n\rangle
 \label{eq:random_state}
\end{equation}
where $|n\rangle$ denotes a computational basis state,
$g_n^{(r)}$ are independent standard normal random variables,
and $\mathcal N_r$ is the normalization factor.
To ensure a fair comparison,
we use the same initial random state
for FTLM, FTiTL, and FTiTK for each sample $r$.
The same initial states are also used for the c-TPQ calculations
at $N=16$ to reduce the effects of statistical fluctuations
arising from differences in the random states.
The spectral width $W$, used to normalize the imaginary-time step,
is determined from the maximum and minimum eigenvalues
estimated using the Lanczos method.

In the following, we first examine the convergence
of thermodynamic quantities with the Krylov dimension $D$
and then quantitatively evaluate the relationship
between the accuracy of the specific heat
and the number of local operator applications.

\subsection{Validation against exact diagonalization}
\label{sec:ed}

We first examine the accuracy of each method
by comparison with exact diagonalization
for the $N=16$ square-lattice transverse-field Ising model
under periodic boundary conditions.
We set the transverse field to $h/J\simeq3.04438$,
corresponding to the quantum critical point
in the thermodynamic limit.
We also perform c-TPQ calculations for comparison.
In c-TPQ, finite-temperature states are constructed
by applying the imaginary-time evolution operator
to initial random states.
In our calculations, the Hamiltonian is normalized
by its spectral width,
and the imaginary-time evolution is approximated
using a Taylor expansion up to order $m$ in $H$.
The specific heat is evaluated from the energy variance,
and the entropy is obtained by thermodynamic integration
of the specific heat.
\cite{Sugiura2013,BritoFerreira2024}

Fig.~\ref{fig:ED} shows the temperature dependence
of the specific heat $C/N$ and entropy $S/N$
for several Krylov dimensions $D$
and c-TPQ expansion orders $m$.
We use $R=100$ random states
and set $\rho=8.0$ and $M=1$ for FTiTL and FTiTK.
We choose $\rho=8.0$ because it provides good convergence,
while $M=1$ is selected to examine convergence
under conditions where the Trotter error is relatively large.

For FTLM, the specific-heat peak and the temperature dependence
of the entropy show substantial deviations from the exact results
at small $D$.
As $D$ increases, both quantities approach the exact values
and are nearly reproduced at $D=25$.
For c-TPQ, substantial deviations are observed
at low temperatures for $m=25$ and $100$.
The accuracy improves with increasing $m$,
and the exact results are nearly reproduced at $m=400$.
In c-TPQ, $m$ corresponds to the number of Hamiltonian applications
per initial random state,
while the Krylov dimension $D$ in FTLM
corresponds to the same number of Hamiltonian applications.
Thus, under the present computational conditions,
FTLM achieves comparable accuracy with fewer
Hamiltonian applications than c-TPQ.
Furthermore, FTiTL and FTiTK reproduce
the exact results closely even at $D=10$,
providing accurate estimates of both thermodynamic quantities
at smaller $D$ than FTLM.
These results demonstrate that,
with an appropriate imaginary-time step,
FTiTL and FTiTK can obtain accurate thermodynamic quantities
with smaller Krylov dimensions than conventional FTLM.

\figpdf[1.0\linewidth]{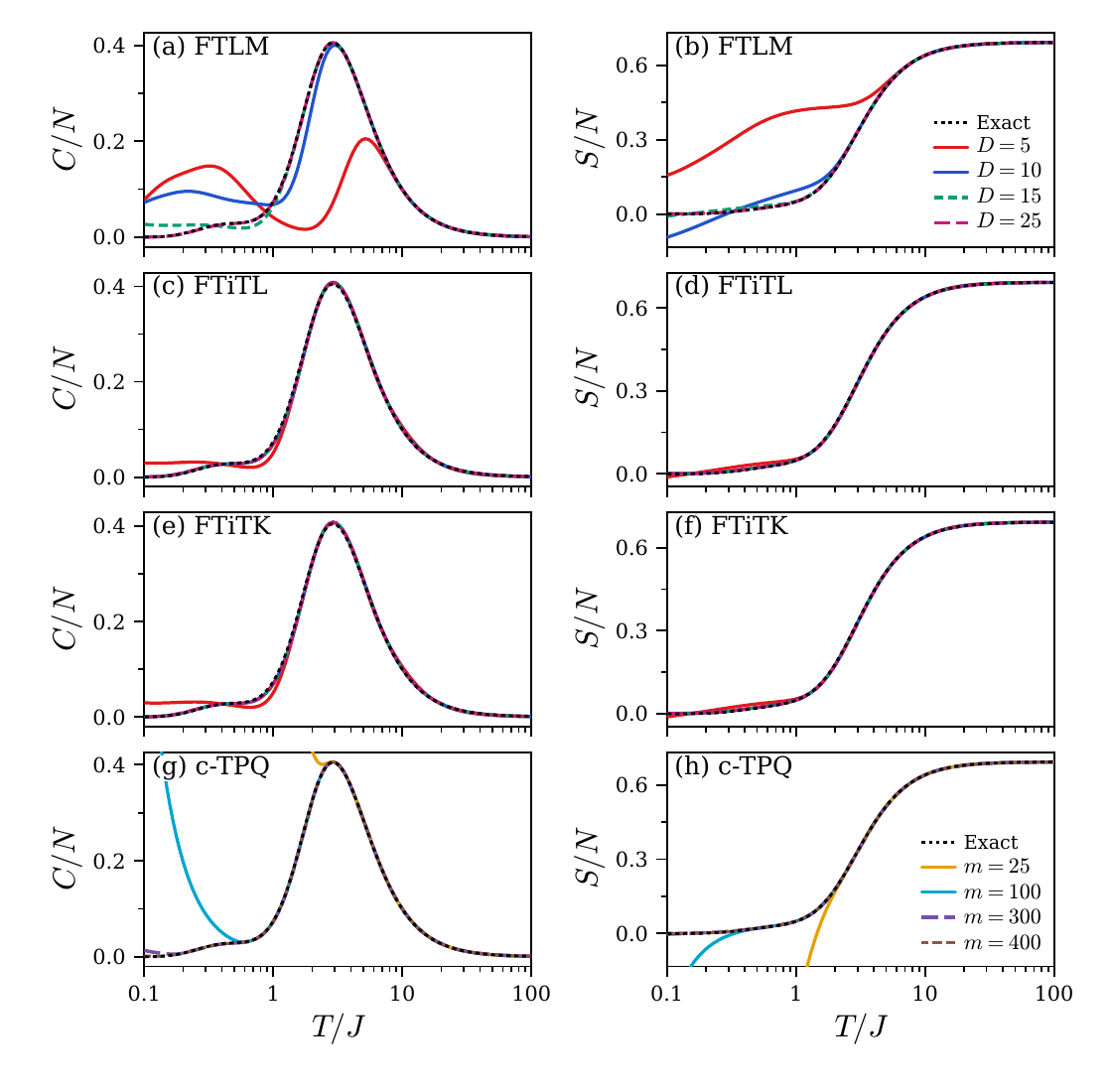}{
(Color online) Comparison of FTLM, FTiTL, FTiTK, and c-TPQ
with exact diagonalization
for the $N=16$ square-lattice transverse-field Ising model.
The left and right columns show the temperature dependence
of the specific heat $C/N$ and entropy $S/N$, respectively.
$J=1$, $h=3.04438$, and $R=100$.
For FTiTL and FTiTK, $\rho=8.0$ and $M=1$ are used.
In c-TPQ, $m$ denotes the number of Hamiltonian applications
per initial random state,
and the Krylov dimension $D$ in FTLM
corresponds to the same number of applications.
\label{fig:ED}
}

\subsection{Thermodynamic properties of the 32-site model}
\label{sec:thermo}

Next, we examine the accuracy of each method
by comparison with QMC calculations
for the $N=32$ square-lattice transverse-field Ising model
under periodic boundary conditions.
We again use the quantum critical field
$h/J\simeq3.04438$,
with $R=100$, $\rho=8.0$, and $M=1$.

Fig.~\ref{fig:QMC} shows the temperature dependence
of the specific heat $C/N$ and entropy $S/N$.
As in the $N=16$ system,
FTLM exhibits substantial deviations from the QMC reference values
at small $D$.
With increasing $D$,
the specific-heat peak and the temperature dependence
of the entropy are better reproduced.
In contrast, FTiTL and FTiTK yield results
closer to the QMC reference values
at smaller Krylov dimensions.
At sufficiently large $D$,
all three methods reproduce the QMC results well.

\figpdf[1.0\linewidth]{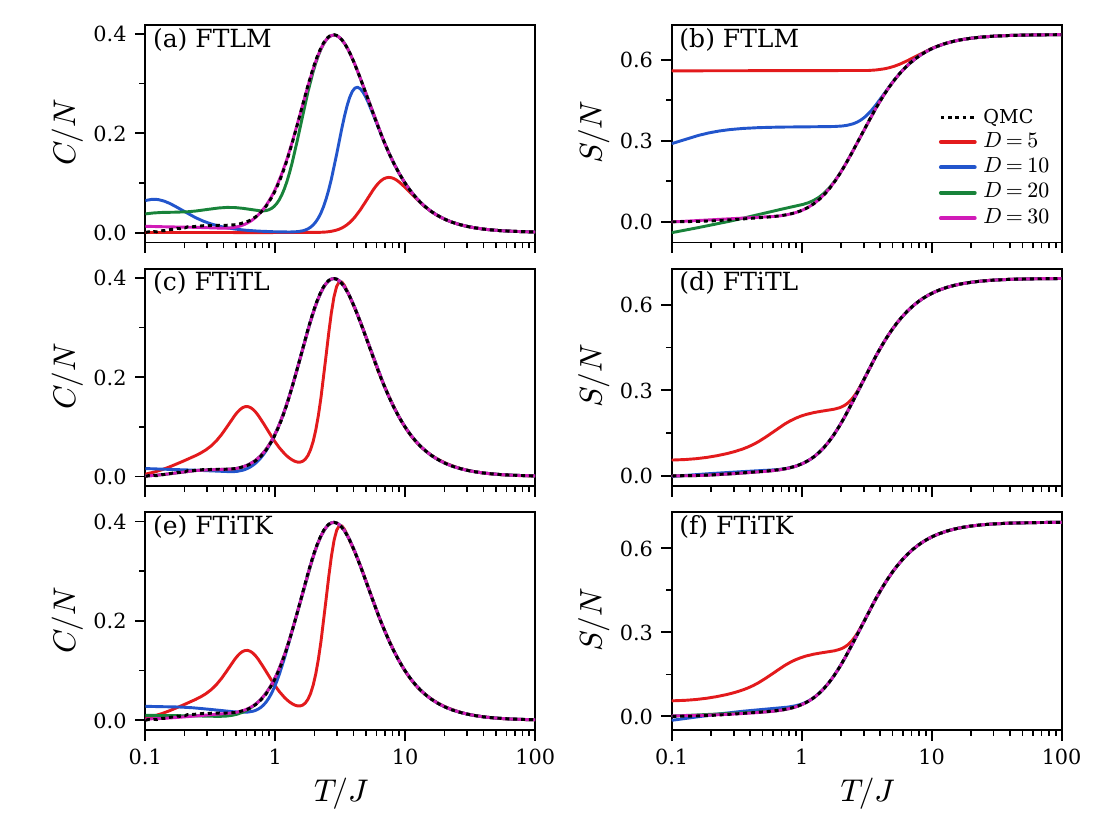}{
\label{fig:QMC}
(Color online) Comparison of FTLM, FTiTL, and FTiTK with QMC
for the $N=32$ square-lattice transverse-field Ising model.
The left and right columns show the temperature dependence
of the specific heat $C/N$ and entropy $S/N$, respectively.
$J=1$, $h=3.04438$, and $R=100$.
For FTiTL and FTiTK, $\rho=8.0$ and $M=1$ are used.
}

These results indicate that the improved convergence
of the imaginary-time Krylov methods,
observed in comparison with exact diagonalization at $N=16$,
is also found in the $N=32$ system.
However, because the computational cost per Krylov dimension $D$
differs among the methods,
a comparison based solely on $D$
is insufficient to assess their relative computational efficiency.
In the next subsection, we quantitatively compare
the accuracy and computational cost
using the number of local operator applications
defined in Sec.~\ref{sec:cost}.

\subsection{Accuracy versus local operator-action counts}
\label{sec:acc_cost}

In this subsection, we compare the accuracy
and computational cost of the three methods.
In Figs.~\ref{fig:ED} and \ref{fig:QMC},
the results of the three Krylov methods
nearly overlap with the reference values as the Krylov dimension $D$ increases,
making it difficult to distinguish their accuracy
through visual inspection alone.
To quantify even small deviations,
we use the root-mean-square error (RMSE)
of the specific heat $C/N$ with respect to the QMC reference values:
\begin{equation}
 \operatorname{RMSE}_{C/N,X}
 =\sqrt{
 \frac{1}{N_T}
 \sum_{i=1}^{N_T}
 \left[
 \frac{C_X(T_i)}{N}-\frac{C_{\mathrm{QMC}}(T_i)}{N}
 \right]^2
 }
 \label{eq:rmse}
\end{equation}
Here, $T_i$ denotes one of the $N_T$ temperature points
used for comparison,
and $C_X$ and $C_{\mathrm{QMC}}$ are the specific heats
obtained by method $X$ and QMC, respectively.
A smaller RMSE indicates better agreement
with the QMC reference values.
To minimize dependence on code implementation
and computing environment,
we use the total number of local operator applications $L_X$,
defined in Sec.~\ref{sec:cost},
as the measure of computational cost.

Fig.~\ref{fig:RMSE} shows the relationship
between the RMSE of the specific heat
and the number of local operator applications $L_X$
for $N=32$.
We use $N_T=150$ logarithmically spaced temperature points
over $0.1\leq T/J\leq100$,
with $R=100$ and $M=1$.
The QMC reference values were computed directly
at all 150 temperature points.

\figpdf[1.0\linewidth]{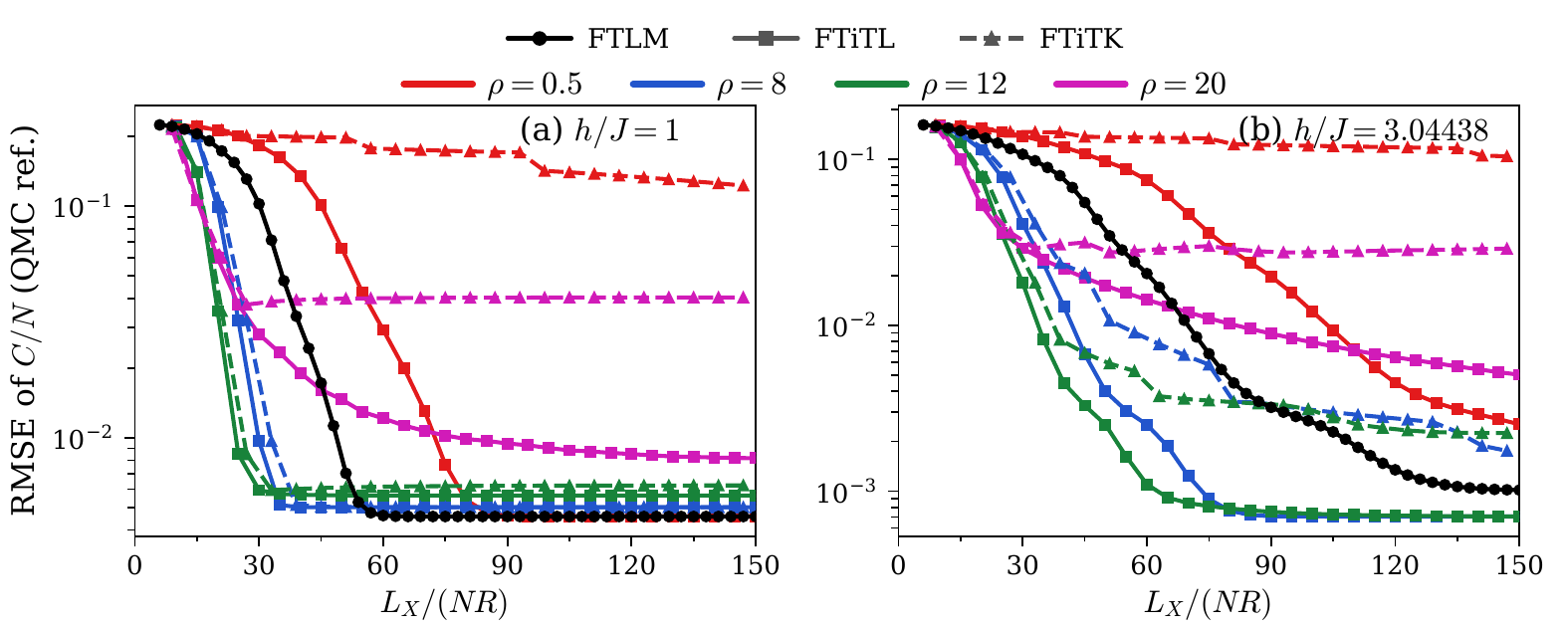}{
\label{fig:RMSE}
(Color online) Relationship between the RMSE of the specific heat
and the number of local operator applications $L_X$
for the $N=32$ square-lattice transverse-field Ising model.
(a) $h=1$ and (b) $h=3.04438$.
$J=1$, $R=100$, and $M=1$.
FTLM is compared with FTiTL and FTiTK
for several imaginary-time step parameters $\rho$.
}

The minimum normalized local operator-action count
$L_X/(NR)$ required to achieve an RMSE in the specific heat $C/N$
below $10^{-2}$ is 51 and 72 for FTLM
at $h/J=1$ and $3.04438$, respectively.
In contrast, FTiTL with $\rho=12$ requires
only 25 and 35, respectively.
These values correspond to a reduction
of approximately 51\% in the number of local operator applications
relative to FTLM for both field strengths.
For FTiTK with the same $\rho=12$,
the corresponding values are 27 and 39,
representing reductions of approximately 47\% and 46\%, respectively.
Thus, with an appropriate choice of $\rho$,
both FTiTL and FTiTK achieve an RMSE
of $10^{-2}$ or less
with approximately half the number of local operator applications
required by FTLM,
both in the ferromagnetic regime at $h=1$
and near the quantum critical point at $h=3.04438$.

On the other hand, the accuracy of the imaginary-time methods
depends strongly on $\rho$.
At $\rho=0.5$, the RMSE decreases slowly,
and particularly for FTiTK,
increasing the Krylov dimension yields little improvement in accuracy.
One possible reason is that, for a small imaginary-time step,
$B=\mathrm{e}^{-\eta H}$ approaches the identity operator,
making the Krylov basis vectors nearly linearly dependent
in numerical calculations.
An error plateau is also observed at $\rho=20$,
possibly due to excessive suppression of high-energy components
and increased Trotter errors associated with a large imaginary-time step.

One possible explanation for the improved convergence
of FTiTL and FTiTK is the spectral transformation
induced by imaginary-time evolution.
Under the transformation $B=\mathrm{e}^{-\eta H}$,
the high-energy part of the spectrum is compressed,
while the eigenvalue spacings in the low-energy part
are relatively enlarged.
This is expected to allow the low-energy spectrum
and excitation gaps to be approximated accurately
with smaller Krylov dimensions.
Although the accuracy of the high-energy eigenvalues
may be relatively degraded,
the contributions of these states to the partition function
are exponentially suppressed by the Boltzmann factors.
Their influence on thermodynamic quantities
is therefore expected to be small.

% Differences in the converged RMSE values at sufficiently large computational costs are also observed between the two field strengths. One possible reason is that, even when the same initial random states are used, the deviations in the estimates arising from the finite number of samples (R) depend on the Hamiltonian.

These results demonstrate that,
for the square-lattice transverse-field Ising model,
FTiTL and FTiTK with appropriately chosen imaginary-time steps
achieve higher accuracy with fewer local operator applications
than conventional FTLM,
thereby improving the computational efficiency
of finite-temperature calculations.

\subsection{Dependence on Trotter decomposition}

To examine the effects of the Trotter approximation
of the imaginary-time evolution operator on numerical accuracy,
we compare FTiTL results
for $M=1,2,4$ Trotter steps.
We use $N=32$, $\rho=8$, and $R=100$,
with the FTLM results at $D=80$
obtained from the same initial random states as FTiTL
serving as reference values.
The FTLM specific heat is sufficiently converged at $D=80$.
The RMSE of the specific heat $C/N$
between the FTLM results at $D=70$ and $D=80$
is approximately $10^{-15}$ and $10^{-9}$
for $h/J=1$ and $3.04438$, respectively.
Both differences are sufficiently small compared with
the error levels shown in Fig.~\ref{fig:trotter}.
Using the same initial random states
reduces the effects of statistical fluctuations
in the comparison.

Fig.~\ref{fig:trotter} shows the relationship
between the RMSE of the specific heat
and the Krylov dimension $D$.
Both in the ferromagnetic regime
and near the quantum critical point,
the RMSE generally decreases with increasing $D$
and converges to an $M$-dependent value
at sufficiently large $D$.
The converged RMSE decreases as $M$ increases.
In particular, doubling $M$ reduces the RMSE
to approximately one-quarter of its original value,
consistent with the $M^{-2}$ error scaling
expected for the second-order Suzuki--Trotter decomposition.

On the other hand, increasing $M$
increases the number of local operator applications
required to apply the imaginary-time evolution operator.
Thus, from the viewpoint of computational efficiency,
there is a trade-off between the accuracy
of the Trotter approximation and computational cost.

\figpdf[1.0\linewidth]{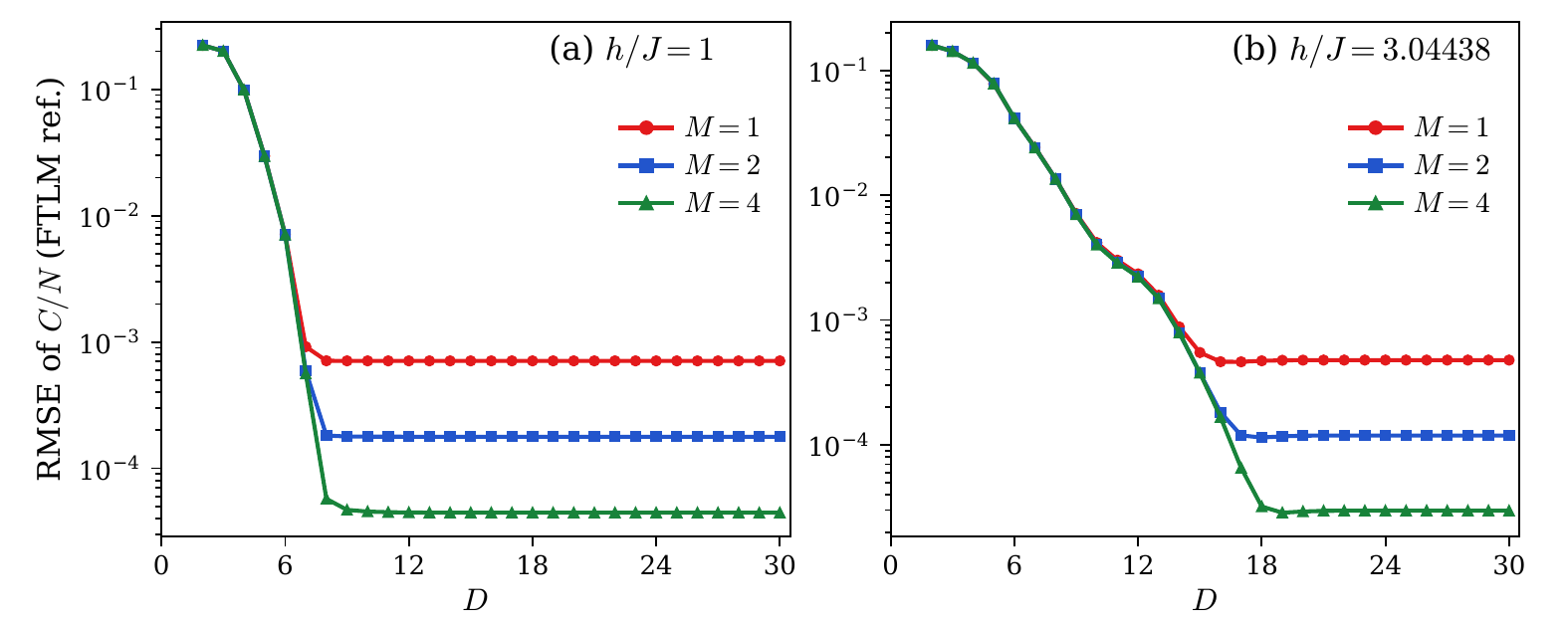}{
\label{fig:trotter}
(Color online) Relationship between the RMSE of the specific heat
and the Krylov dimension $D$
for the $N=32$ square-lattice transverse-field Ising model.
The FTLM results at $D=80$,
obtained using the same initial random states as FTiTL,
are used as reference values.
The RMSE is calculated using Eq.~(\ref{eq:rmse}),
with the QMC reference values replaced
by the FTLM results at $D=80$.
(a) $h=1$ and (b) $h=3.04438$.
$J=1$, $R=100$, and $\rho=8.0$.
FTiTL results for $M=1,2,4$ Trotter steps are compared.
}

\section{Discussion}\label{sec:discussion}

We first compare the computational cost of the proposed methods
with that of c-TPQ.
In c-TPQ, the expansion order $m$ corresponds to the number
of Hamiltonian applications per random state,
which dominate its computational cost.
As shown in Fig.~\ref{fig:ED},
c-TPQ at $m=400$ and FTiTL at $D=10$
both nearly reproduce the exact specific heat and entropy
for $N=16$.
The corresponding normalized numbers of local operator applications,
$L/(NR)$, are $1200$ and $50$, respectively.
Thus, in this comparison, FTiTL requires only $1/24$
as many local operator applications as c-TPQ.

Computational efficiency must also be considered alongside
memory requirements and numerical stability.
In our implementations, FTLM, FTiTL, and FTiTK require
two, three, and one state vectors of full Hilbert-space
dimension, respectively,
while c-TPQ can be implemented using two state vectors.
FTiTL and FTiTK use the same Krylov subspace
and yield identical Ritz information in exact arithmetic.
However, the overlap matrix in FTiTK can become ill-conditioned
due to the nonorthogonal basis.
Thus, FTiTL is a useful choice when computational efficiency
and numerical stability are prioritized,
whereas FTiTK is advantageous when reducing memory requirements
is the primary concern.

The choice of the normalized imaginary-time step parameter
$\rho$ involves a trade-off between convergence
and numerical stability.
For small $\rho$, $B$ approaches the identity operator,
and the nonorthogonal basis in FTiTK
can become nearly linearly dependent.
For large $\rho$, the Trotter error may increase
if the number of Trotter steps $M$ is kept fixed,
whereas increasing $M$ increases the computational cost.
In our approach, the imaginary-time step is normalized
as $\eta=\rho/W$.
For many systems with short-range interactions,
including the present model,
the spectral width $W$ grows proportionally
with the system size $N$.
Thus, at fixed $\rho$ and $M$, $\eta$ decreases
with increasing system size,
which is expected to suppress local errors
associated with the Trotter approximation.

For $N=32$, good convergence was obtained
with $\rho=8$--$12$ at both $h/J=1$ and $3.04438$.
Good convergence was also confirmed at $\rho=8$
for $N=16$.
These results suggest that suitable values of $\rho$
for larger systems may be estimated
from preliminary calculations on smaller systems.

\section{Conclusions}\label{sec:conclusions}

We developed two finite-temperature Krylov methods,
FTiTL and FTiTK, based on imaginary-time evolution
and evaluated their accuracy and computational cost
for the square-lattice transverse-field Ising model.
Using exact diagonalization for $N=16$
and quantum Monte Carlo results for $N=32$ as references,
we demonstrated that both methods can achieve
higher computational efficiency than conventional FTLM
with appropriately chosen imaginary-time steps.

For $N=32$, FTiTL reduced the number of local operator
applications required to achieve an RMSE of $10^{-2}$ or less
in the specific heat $C/N$
by approximately 50\% relative to FTLM.
FTiTK also reduced the computational cost
and required only one state vector
of full Hilbert-space dimension,
halving the memory required for storing state vectors
relative to FTLM.

\acknowledgments
The authors thank N.~Shibata for valuable discussions.
This work was supported by JSPS KAKENHI Grant No.~JP25K07230.
The numerical calculations were performed using the
computational facilities of the Supercomputer Center,
Institute for Solid State Physics, The University of Tokyo.
ChatGPT (OpenAI) was used to assist with numerical code development
and manuscript revision.

\end{document}